\documentclass[runningheads]{llncs}
\usepackage{graphicx}
\usepackage{subfig}

\usepackage{color}
\definecolor{blue(pigment)}{rgb}{0.2, 0.2, 0.6}

\def\blue#1{{\color{blue}#1}}

\usepackage{lineno,hyperref}
\modulolinenumbers[5]

\usepackage{hyperref}
\hypersetup{
colorlinks=true,
citecolor=blue,
linkcolor=blue,
filecolor=magenta,
urlcolor=blue
}
\usepackage{colortbl}%
  
\usepackage{xspace}
\usepackage{amssymb}
\usepackage{pifont}
\usepackage{footnote}
\usepackage{multicol}
\usepackage{multirow}
\usepackage{booktabs}
\usepackage{array,makecell}
\usepackage[export]{adjustbox}
\usepackage{arydshln}
\usepackage{graphicx}
\usepackage{subfig}
\usepackage{rotating}
\usepackage{wasysym}
\usepackage{soul}
\usepackage{xcolor}
\usepackage{color}
\usepackage{wrapfig}
\usepackage{caption}

\newcommand{\gow}{{Gowalla}\xspace}
\newcommand{\yelp}{{Yelp}\xspace}

\usepackage{pgfplots}
\pgfplotsset{width=10cm,compat=1.14}
\usepackage{tikz,pgfplotstable}
\usepackage[utf8]{inputenc}
\usepackage{pgfplots,pgfplotstable,booktabs}
\usetikzlibrary{positioning}

\newcommand{\goso}{\texttt{GeoSoCa}\xspace}

\newcommand{\lore}{\texttt{LORE}\xspace}
\newcommand{\topp}{\texttt{MostPop}\xspace}

\newcommand{\bpr}{\texttt{BPR}\xspace}
\newcommand{\wmf}{\texttt{WMF}\xspace}
\newcommand{\pf}{\texttt{PF}\xspace}
\newcommand{\vaecf}{\texttt{VAECF}\xspace}
\newcommand{\neumf}{\texttt{NeuMF}\xspace}

\newcommand{\eg}{e.g., }
\newcommand{\ie}{i.e., }
\newcommand{\vs}{vs.~}
\newcommand{\wrt}{w.r.t. }

\begin{document}
\title{The Unfairness of Active Users and Popularity Bias in Point-of-Interest Recommendation}
\titlerunning{The Unfairness in Point-of-Interest Recommendation}
%

\author{Hossein A.~Rahmani\inst{1}\thanks{Corresponding author.} \and
Yashar Deldjoo\inst{2} \and
Ali Tourani\inst{3} \and \\
Mohammadmehdi Naghiaei\inst{4}}

\authorrunning{H.~A.~Rahmani, Y.~Deldjoo, A.~Tourani, M.~Naghiaei}

\institute{
University College London, United Kingdom \\ \email{h.rahmani@ucl.ac.uk} \and
Polytechnic University of Bari, Italy \\ \email{deldjooy@acm.org} \and
University of Guilan, Iran \\ \email{tourani@msc.guilan.ac.ir} \and
University of Southern California, USA \\ \email{naghiaei@usc.edu}}

\maketitle              
\begin{abstract}
Point-of-Interest (POI) recommender systems provide personalized recommendations to users and help businesses attract potential customers. Despite their success, recent studies suggest that highly data-driven recommendations could be impacted by data biases, resulting in unfair outcomes for different stakeholders, mainly consumers (users) and providers (items). Most existing fairness-related research works in recommender systems treat user fairness and item fairness issues individually, disregarding that RS work in a two-sided marketplace. This paper studies the interplay between \textit{(i)} the unfairness of active users, \textit{(ii)} the unfairness of popular items, and \textit{(iii)} the accuracy (personalization) of recommendation as three angles of our study triangle.

We group users into advantaged and disadvantaged levels to measure user fairness based on their activity level. For item fairness, we divide items into short-head, mid-tail, and long-tail groups and study the exposure of these item groups into the top-$k$ recommendation list of users. 
Experimental validation of eight different recommendation models commonly used for POI recommendation (e.g., contextual, CF) on two publicly available POI recommendation datasets, \gow and \yelp, indicate that most well-performing models suffer seriously from the unfairness of popularity bias (provider unfairness).  Furthermore, our study shows that most recommendation models cannot satisfy both consumer and producer fairness, indicating a trade-off between these variables possibly due to natural biases in data. We choose the POI recommendation as our test scenario; however, the insights should be trivially extendable on other domains.   
\end{abstract}

\section{Introduction}
\label{sec:intro}
Point-of-Interest (POI) recommendation is an essential service to location-based social networks (LBSNs). Providing personalized POI systems ease the inevitable problem of information overload on users and helps businesses attract potential customers~\cite{sanchez2021point,rahmani2020joint}. However, as a highly data-driven system, these systems could be impacted by data or algorithmic bias, providing unfair outcomes and weakening the system's trustworthiness. Consequently, bias and fairness have attracted rapidly growing attention in machine learning and recommender system research communities~\cite{deldjoo2021flexible,chen2020bias,mehrabi2021survey}. 

Traditional recommender systems (RS) focused on maximizing customer satisfaction by tailoring the content according to individual customers' preferences, thereby disregarding the interest of the producers. Several recent studies \cite{li2021user,abdollahpouri2019unfairness} have revealed how such customer-centric designs may impair the well-being of the producers. Thus, in recent years, two-sided marketplaces have steadily emerged as a pivotal problem of fairness topics in RS, requiring optimizing supplier preferences and visibility.

Most papers in this field focus on the fairness perspective either on the side of users or items~\cite{gomez2022provider,boratto2022consumer,abdollahpouri2019unfairness,kowald2020unfairness,abdollahpouri2021user}. Contrary to these works, the work at hand aims to look into the interplay between \textit{accuracy}, \textit{producer fairness}, and \textit{consumer fairness} in a multi-sided marketplace ecosystem and the possible trade-off between them. Specifically, we focus on the fairness of state-of-the-art POI recommendation algorithms and investigate the performance of various algorithms on two real-world POI datasets, \gow and \yelp, to address the following research questions:




\begin{itemize}
    \item \textbf{RQ1}: To what extent are users interested in popular POIs? That is identifying natural data bias toward popularity. (cf.~Section \ref{sec:profile_analysis})
    \item \textbf{RQ2}: Is there a trade-off among the factors of accuracy, user fairness, and item fairness? (cf.~Section \ref{sec:tradeoff_analysis})
    \item \textbf{RQ3}: Could we classify algorithms based on their performance on the accuracy, user fairness, and item fairness? Which ones can produce satisfactory results on all three factors, and which one suffers more from bias issues? (cf.~Section \ref{sec:biasrecommendation})
\end{itemize}

To answer the mentioned questions, we analyze the users' profiles regarding their check-ins behavior and compare several state-of-the-art recommendation algorithms regarding their general popularity bias propagation and the extent to which it affects different groups of users and items. 
In the following, we review the related works in Section \ref{sec:relatedworks} and explain the experimental setup in Section \ref{sec:experiment}. Then, we conduct experiments and the evaluation results in Section \ref{sec:results}. We conclude the paper in Section \ref{sec:conclusion}. Finally, to enable reproducibility of the results, we have made our codes open source.\footnote{\url{https://github.com/RecSys-lab/FairPOI}}

\section{Related Work}
\label{sec:relatedworks}
Fairness is becoming one of the most influential topics in recommender systems in recent years~\cite{chen2020bias,olteanu2021facts,li2021user}. Burke et al.~\cite{burke2017multisided} classified fairness in the recommendation system based on general beneficiaries, consumers (C), providers (P), and both (CP). Deldjoo et al.~\cite{deldjoo2021flexible} proposed a flexible framework to evaluate consumer and provider fairness using generalized cross-entropy. 
Li et al.~\cite{li2021user} focused on the fairness in the recommendation systems from the user's perspective in the e-commerce domain, \ie C-fairness. They created user groups based on their activity level into two groups: advantaged and disadvantaged. They showed that users in the advantaged group (\ie active users) usually receive higher quality recommendations than disadvantaged groups, while the advantaged users are a small fraction of all users. Then, they proposed a fair re-ranking model by introducing constraints based on the 0-1 integer programming to reduce this gap. Abdollahpouri et al.~\cite{abdollahpouri2021user} further researched user-centered evaluation of popularity bias. They proposed a user-centered evaluation method that can effectively tackle popularity bias for different user groups while accounting for users' tolerance towards popularity bias using Jensen divergence. Similarly, we divide the users into four groups based on their activity level and calculate the recommendation accuracy and fairness achieved by state-of-the-art algorithms in the location-based recommender systems among user groups. 

Kowald et al.~\cite{kowald2020unfairness} attempted to investigate the user fairness in music recommendation on the Last.fm dataset. The analysis of this paper is in line with the investigation of Abdollahpouri et al.~\cite{abdollahpouri2019unfairness} in the domain of movie recommendation. To do this, they split users into three groups and explored the correlation between users' profiles and well-known artists. The result further indicated that the low-mainstream user groups receive a low recommendation quality in all cases.
Zhang et al.~\cite{zhang2021causal} examined adjusting item popularity bias in recommendation score using the causal intervention and latent factor model by experimenting on three real-world datasets. Similar to the work presented by Abdollahpouri et al.~\cite{abdollahpouri2019unfairness}, we investigate popularity bias in the POI domain on provider perspectives, \ie P-fairness, by dividing them into three categories: \textit{short-head}, \textit{mid-tail}, and \textit{long-tail} locations. We then explore the item exposure fairness of these three distinct categories, thus P-Fairness.

Weydemann et al.~\cite{weydemann2019defining} investigated fairness in the location-based recommender systems by defining various criteria for measuring fairness. Their experiment showed that unfairness in location recommender systems is likely related to location popularity or inclination toward some user groups based on nationality. Lesota et al.~\cite{lesota2021analyzing} investigated item popularity fairness in the music domain using the LFM-2b\footnote{\url{http://www.cp.jku.at/datasets/LFM-2b/}} dataset, \ie P-fairness, and addressed shortcomings of the statistical analysis in fairness by considering the distribution between user-profiles and recommendation lists. Furthermore, they analyzed whether such algorithmic popularity bias affects users of different genders. 

Contrary to the works mentioned above that focus on consumers~\cite{li2021user,abdollahpouri2021user,abdollahpouri2019unfairness}, or provider fairness~\cite{abdollahpouri2019managing,kowald2020unfairness,zhang2021causal}, our work studies the fairness of recommended items from both perspectives (i.e., CP-Fairness). Mehrotra et al.~\cite{mehrotra2018towards} showed that blindly optimizing for consumer relevance might hurt supplier fairness. Hence, one of the main benefits of this study is to allow us to understand whether there exists a possible trade-off between these factors $<$accuracy, user fairness, provider fairness$>$ and measure how much we need to sacrifice consumer satisfaction in terms of relevance to have a more fair marketplace. 


\section{Experimental Setup}
\label{sec:experiment}

\subsection{Datasets}
In this work, we use two well-known check-in datasets, namely, \gow and \yelp provided by~\cite{liu2017experimental}\footnote{\url{http://spatialkeyword.sce.ntu.edu.sg/eval-vldb17/}}. The \gow dataset was collected from February 2009 to October 2010. Following~\cite{liu2017experimental,rahmani2020lglmf,baral2018exploiting}, we preprocessed the \gow dataset by removing cold users, \ie users who visited locations for less than $15$ check-ins. We also excluded POIs of less than $10$, which may cause a spam error on the model. The \yelp dataset is provided by the \yelp Dataset Challenge\footnote{\url{https://www.yelp.com/dataset/challenge}} round 7 (access date: Feb 2016) in 10 metropolitan areas across two countries. Also, in this case, we preprocessed the dataset and removed users who had less than $10$ visited locations and  POIs with less than $10$ visits. Table~\ref{tbl:datasets} shows the statistics of the final datasets after prepossessing steps.

\begin{table}[t]
  \caption{Characteristics of the datasets used in the evaluation: $\left| \mathcal{U} \right|$ is the number of users, $\left| \mathcal{P} \right|$ is the number of POIs, $\left| \mathcal{C} \right|$ is the number of check-ins, $\left| \mathcal{S} \right|$ the number of social link, $\left| \mathcal{G} \right|$ is the number of categories (\eg restaurant, cafe), $\frac{\left| \mathcal{C} \right|}{\left| \mathcal{U\times{P}} \right|}$ is the density. $-$ shows that on the \gow dataset we do not have categorical information of POIs.}
  \centering
  \label{tbl:datasets}
  \begin{tabular}{ccccccccc}
    \toprule
    \textbf{Dataset} & $\left| \mathcal{U} \right|$ & $\left| \mathcal{P} \right|$ & $\left| \mathcal{C} \right|$ & $\left| \mathcal{S} \right|$ & $\left| \mathcal{G} \right|$ & $\frac{\left| \mathcal{C} \right|}{\left| \mathcal{U} \right|}$ & $\frac{\left| \mathcal{C} \right|}{\left| \mathcal{P} \right|}$ & $\frac{\left| \mathcal{C} \right|}{\left| \mathcal{U\times{P}} \right|}$ \\
    \midrule
    \multirow{1}{*}{\textbf{\yelp}} 
    & 7,135 & 15,575 & 299,327 & 46,778 & 582 & 41.95 & 19.21 & 0.0026 \\
    \midrule
    \multirow{1}{*}{\textbf{\gow}} 
    & 5,628 & 30,943 & 618,621 & 46,001 & $-$ & 109.91 & 19.99 & 0.0035 \\
    \bottomrule 
  \end{tabular}
\end{table}

\subsection{Evaluation Metrics}
For evaluation, we use NDCG, and two fairness evaluation metrics, namely, Generalized Cross-Entropy (GCE) and Mean Absolute Deviation of ranking performance (MADr)~\cite{deldjoo2021flexible}. Both metrics can be used to measure user and item fairness. The difference is that GCE compares the recommendation distribution with a fair one, while the latter compares the absolute deviation between recommendation models. If the attribute $a \in A$ is discrete or categorical, then the unfairness measure according to GCE is defined as:

\begin{equation}
\label{eq:GCE_discrete}
    GCE(m, a) = \frac{1}{\beta \cdot (1-\beta)}\left[\sum_{a_j} p_f^\beta(a_j) \cdot p_m^{(1-\beta)}(a_j)  -1\right]
\end{equation}

\noindent
in which $p_m$ and $p_f$ stand for recommendation model and target distributions, respectively. Note that the defined unfairness measure indexed by $\beta$ includes the Hellinger distance for $\beta=1/2$, the Pearson's $\chi^2$ discrepancy measure for $\beta=2$, Neymann's $\chi^2$ measure for $\beta=-1$, the Kullback-Leibler divergence in the limit as $\beta \rightarrow 1$, and the Burg CE distance as $\beta \rightarrow 0$. MAD for ranking (MADr) can be defined formally by:

\begin{equation}
   MADr(i, j) = \left | rank^{(i)}  - rank^{(j)} \right |
\end{equation}

\noindent
where $rank^{(i)}$ denotes the average ranking performance restricted to those users in group $i$, and $rank^{(j)}$ captures the same metric score for group $j$. The reported MADr corresponds to the average MADr between all the pairwise combinations within the groups involved, \ie $MADr = \mbox{avg}_{i,j}{(MADr(R^{(i)}, R^{(j)}))}$. Larger values for MADr imply differentiation between groups interpreted as unfairness. 

To evaluate the performance of the recommendation methods we partition each dataset into training, validation, and test data. For each user, we use the earliest 70\% check-ins as training data, the most recent 20\% check-ins as test data, and the remaining 10\% as validation data.

\begin{figure}[!tbp]
\captionsetup[subfloat]{font=tiny}
  \centering
    \subfloat[Long-tail of check-ins counts]{\includegraphics[width=0.25\textwidth]{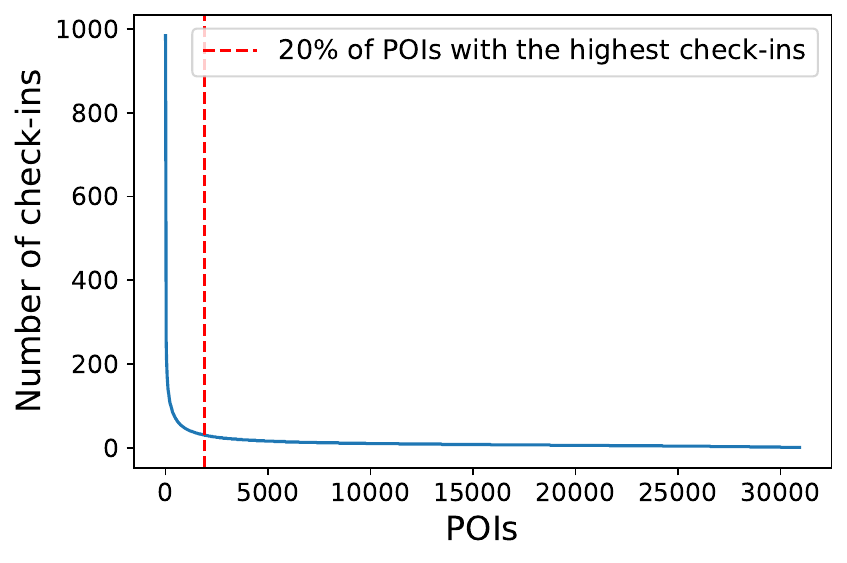}\label{fig:gowalla_data_analysis_1}}
    \hfill
    \subfloat[Popular POIs in user profiles]{\includegraphics[width=0.25\textwidth]{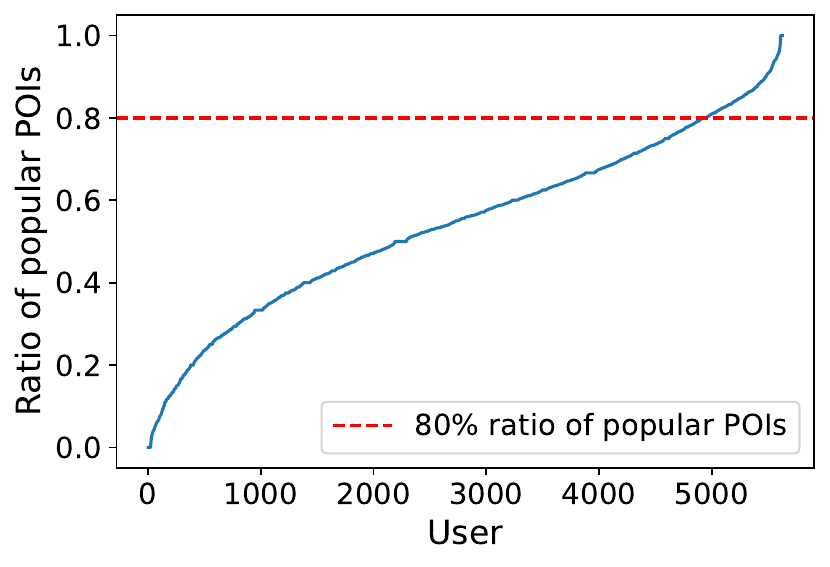}\label{fig:gowalla_data_analysis_2}}
    \hfill
    \subfloat[Number of popular POIs]{\includegraphics[width=0.25\textwidth]{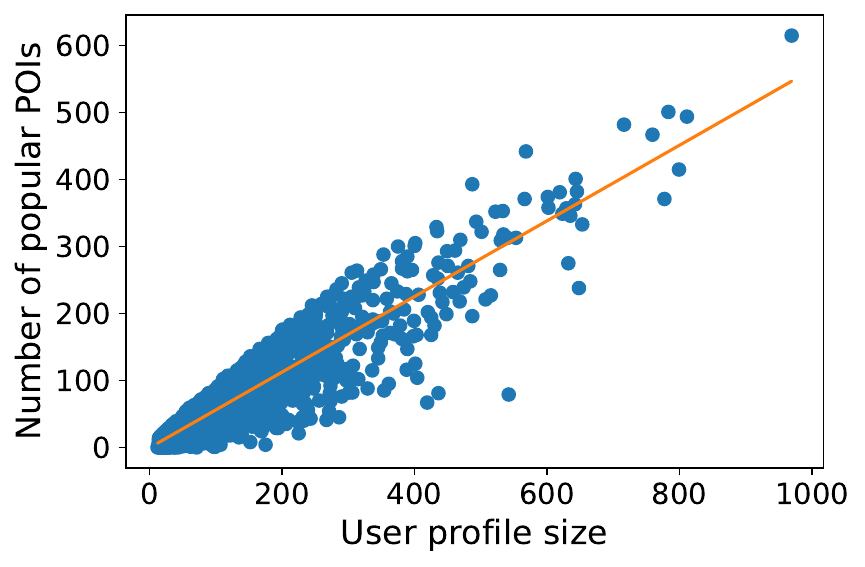}\label{fig:gowalla_data_analysis_3}}
    \hfill
    \subfloat[Average popularity of POIs]{\includegraphics[width=0.25\textwidth]{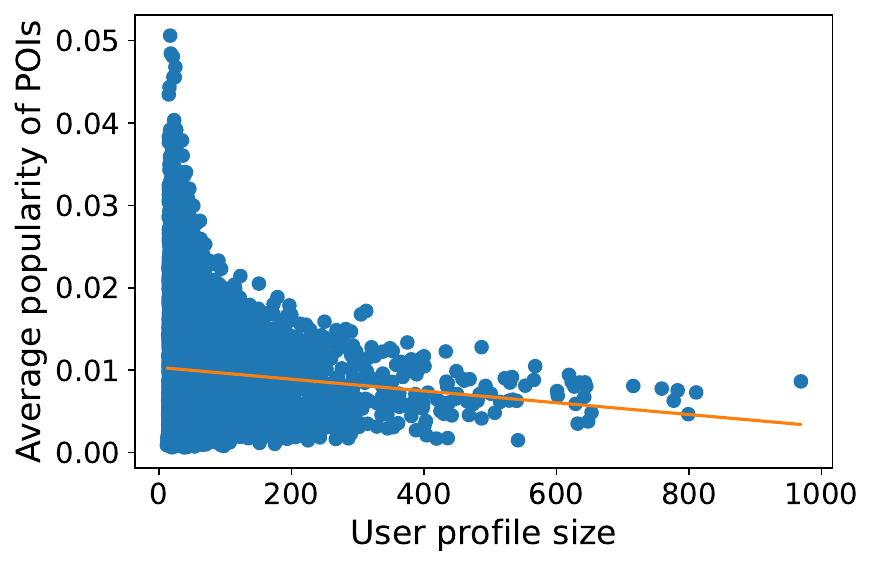}\label{fig:gowalla_data_analysis_4}}
    \hfill
    \subfloat[Long-tail of check-ins counts]{\includegraphics[width=0.25\textwidth]{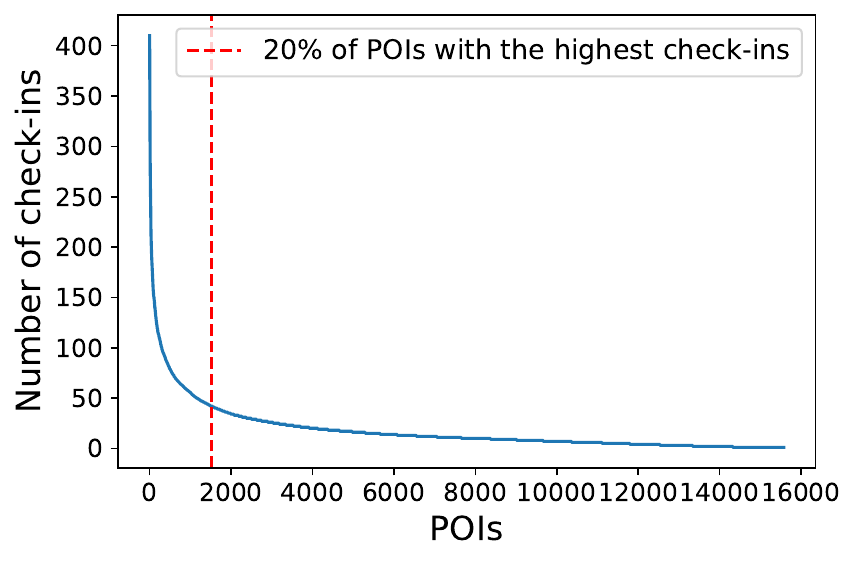}\label{fig:yelp_data_analysis_1}}
    \hfill
    \subfloat[Popular POIs in user profiles]{\includegraphics[width=0.25\textwidth]{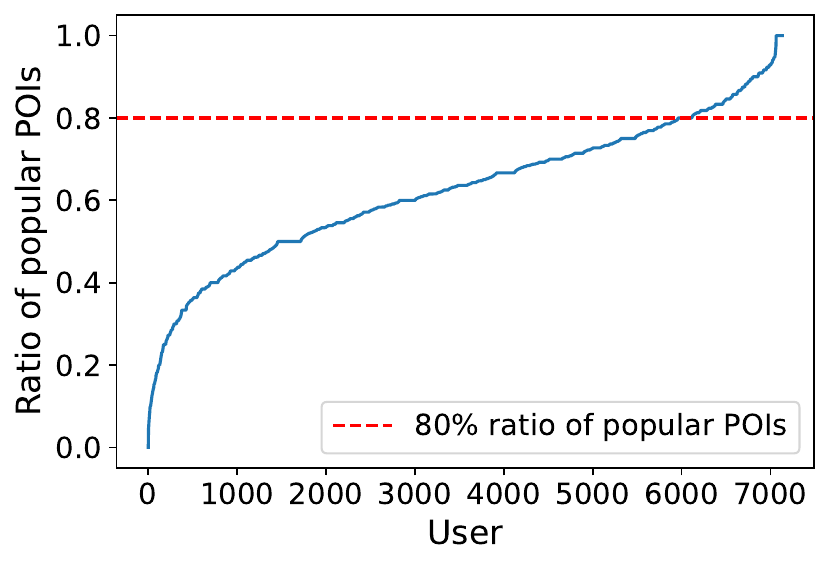}\label{fig:yelp_data_analysis_2}}
    \hfill
    \subfloat[Number of popular POIs]{\includegraphics[width=0.25\textwidth]{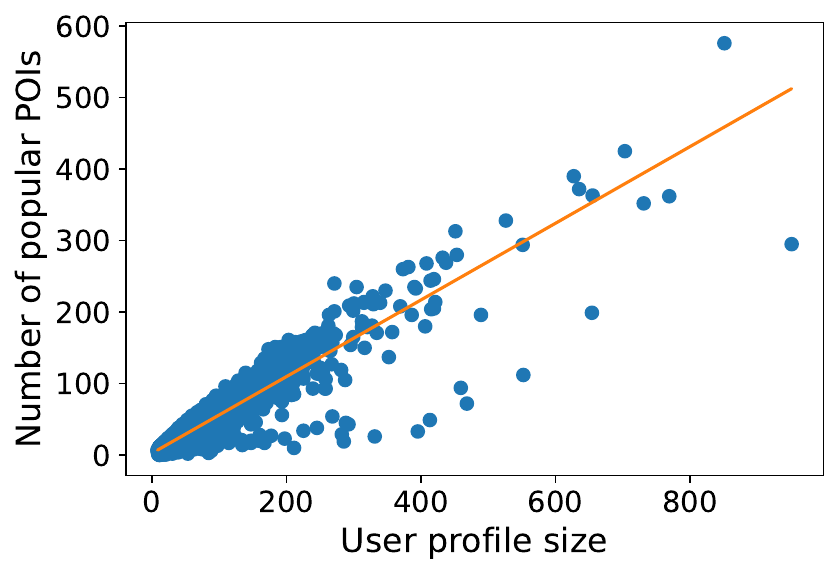}\label{fig:yelp_data_analysis_3}}
    \hfill
    \subfloat[Average popularity of POIs]{\includegraphics[width=0.25\textwidth]{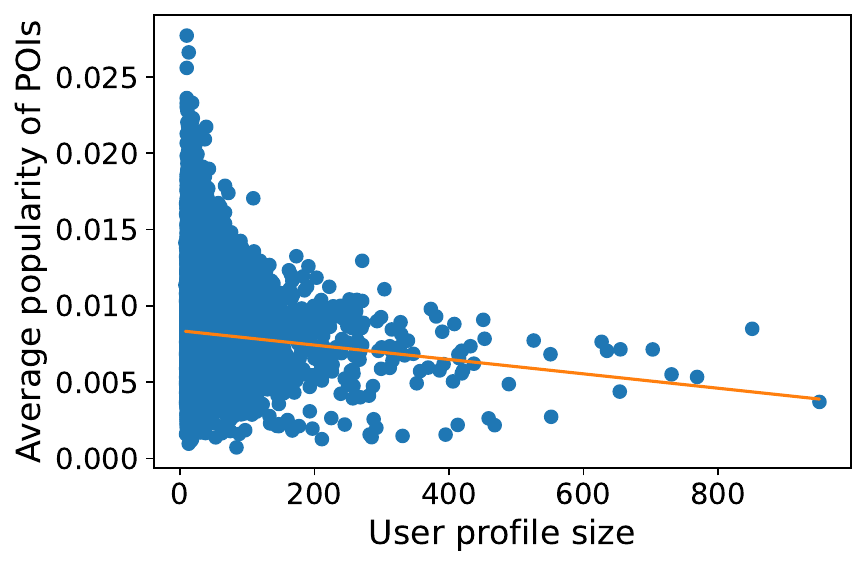}\label{fig:yelp_data_analysis_4}}
  \caption{Figs.~(a-d) and (e-h) show evaluations on \gow and \yelp datasets, respectively. Figs.~(a), (b), (e), and (h) are long-tail of check-ins, and (c), (d), (g), (h) refer to the correlation of user profile size and the popularity of artists in the user profile.}
  \label{fig:data_analysis}
\end{figure}

\section{Popularity Bias in POI Data (RQ1)}
\label{sec:profile_analysis}
In this section, to answer the first research question, we analyze the consumption distribution of POIs across two datasets and the degree of interest that different types of users have towards popular items.

\subsection{Consumption Distribution of POIs}
In Fig \ref{fig:data_analysis}, we show the consumption distribution of POIs across the two datasets chosen for this work, namely, \gow and \yelp. In both Figs.~\ref{fig:gowalla_data_analysis_1} and \ref{fig:yelp_data_analysis_1}, we observe a long-term distribution of the POI checks-ins, where few items (POIs) are consumed (visited) by many users, while few users only see most POIs.
We highlight the popular items in both datasets by segmenting each into three categories (\textit{short-head}, \textit{mid-tail}, and \textit{long-tail}) in such a way that \textit{short-head} items correspond to 50\% of check-ins, while \textit{mid-} and \textit{long-tail} items provide the remaining 30\% and 20\%, respectively.  Furthermore, following \cite{liu2017experimental}, user groups have been divided into \textit{very inactive}, \textit{slightly inactive}, \textit{slightly active}, and \textit{very active} based on their number of check-ins. Users of \gow were mapped to the mentioned classes as ``$<$19'', ``19-47'', ``47-94'', and ``$>$94'', containing $921$, $1992$, $1266$, and $1449$ instances, respectively. The same pattern goes for \yelp with divisions ``$<$51'', ``51-128'', ``128-256'', and ``$>$256'', each with $3099$, $2444$, $954$, and $638$ users.

Notably, it can be seen that on \yelp, $9.74\%$ of items have a higher number of check-ins (1,517 items out of 15,575), while the same statistic for \gow is 6.19\%  (1,914 items out of 30,943). Accordingly, it should be noted that the popular items constitute a small portion (less than 10\%) of the catalog. In addition, in Figs.~\ref{fig:gowalla_data_analysis_2} and \ref{fig:yelp_data_analysis_2}, we plot the popular items in the user profiles for both datasets. We see that for \gow 4,910 out of 5,628 users (\ie around $87\%$ of users) have at least 20\% unpopular items in their profiles, while this number for \yelp corresponds to 6,116 out of 7,135 users (\ie about $86\%$). 

\subsection{User Profiles and Popularity Bias}
Figs.~\ref{fig:gowalla_data_analysis_3} and \ref{fig:yelp_data_analysis_3} discuss the correlation between the popularity of items in the user profile and the user profile size on both datasets. We can see that the correlation between the number of popular POIs in the user profile over the profile size is positive, where the values of correlation coefficient \textit{R} are $0.9319$ and $0.9355$ for \gow and \yelp, respectively. It can be understood that increasing the number of items in a user profile extends the probability of finding popular POIs. In contrast, the correlation between the average popularity of POIs over the user profile sizes is negative (\textit{R} = -0.082 for \gow and \textit{R} = -0.0785 for \yelp). Thus, users with fewer items included in their profiles prefer to attend more popular POIs.

\section{Results and Discussion}
\label{sec:results}
This section studies active users and popularity bias in state-of-the-art POI recommendation algorithms. To foster the reproducibility of our study, we calculate and evaluate all recommendations with the Python-based open-source recommendation toolkit Cornac\footnote{\url{https://cornac.preferred.ai/}}. Using Cornac, we formulate our POI recommendations as a top-$k$ recommendation problem, where we recommend a target user a list of POI items, including top-10 POI items with the highest preferences.

We use different types of state-of-the-art algorithms to evaluate and compare, which
includes (i) conventional approaches (\topp, \bpr \cite{rendle2009bpr}, \wmf \cite{hu2008collaborative}, and \pf \cite{gopalan2015scalable}), (ii) neural CF approaches (\vaecf \cite{liang2018variational} and \neumf \cite{he2017neural}), and (iii) contextual-based POI recommendation approaches (\ie \goso\footnote{We evaluate \goso only on \yelp as we do not have access to the categorical data of the \gow dataset.} \cite{zhang2015geosoca} and \lore \cite{zhang2014lore}).

\begin{figure}[!tbp]
    \centering
    \subfloat[Acc.~v.s.~user fairness.]{\includegraphics[width=0.33\textwidth]{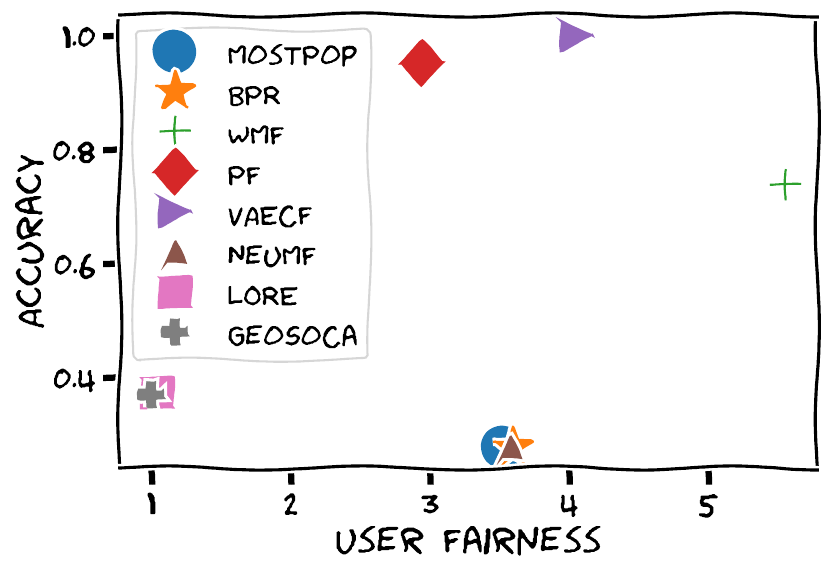}\label{fig:yelp_au}}
    \hfill
    \subfloat[Acc.~\vs item fairness.]{\includegraphics[width=0.33\textwidth]{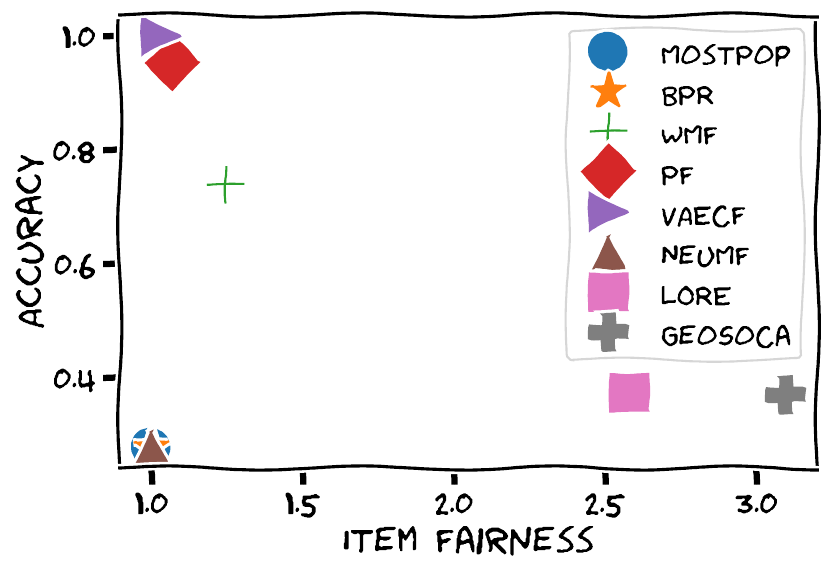}\label{fig:yelp_ai}}
    \hfill
    \subfloat[Item \vs user fairness.]{\includegraphics[width=0.33\textwidth]{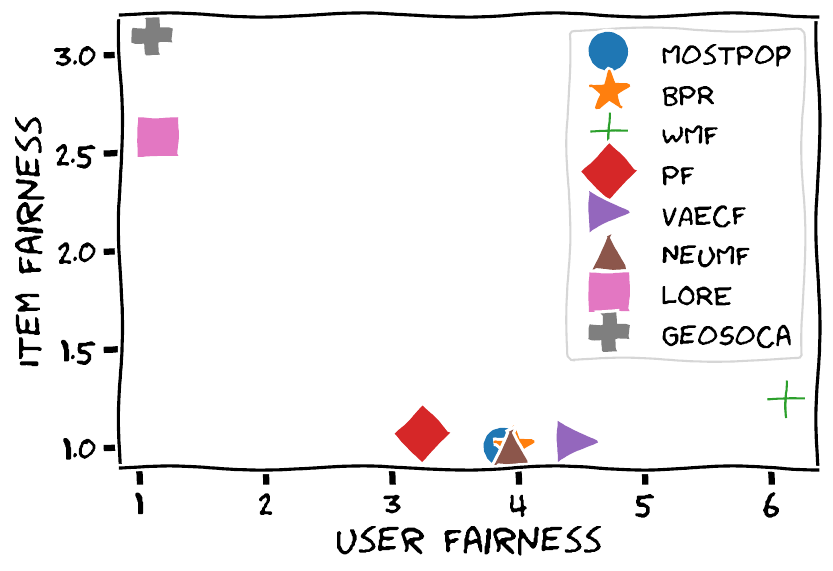}\label{fig:yelp_ui}}
    \hfill
    \subfloat[Acc.~\vs user fairness.]{\includegraphics[width=0.33\textwidth]{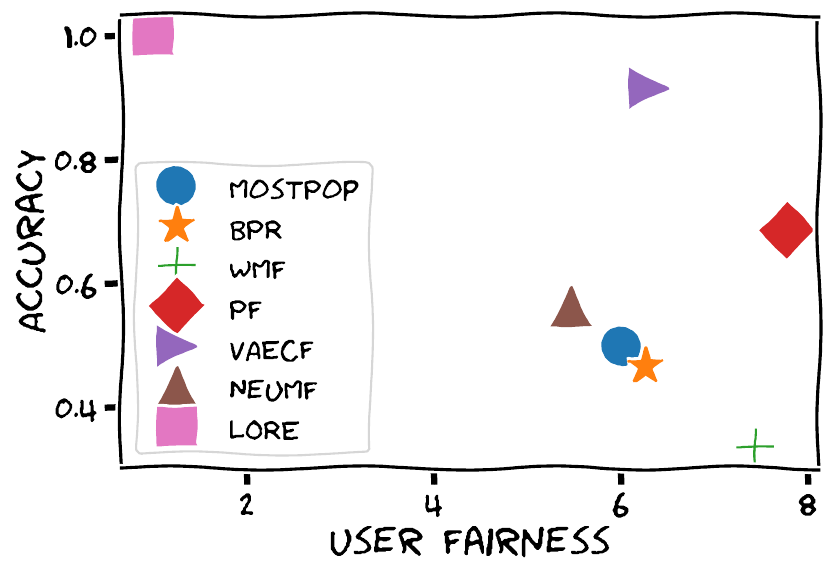}\label{fig:gowalla_au}}
    \hfill
    \subfloat[Acc.~\vs item fairness.]{\includegraphics[width=0.33\textwidth]{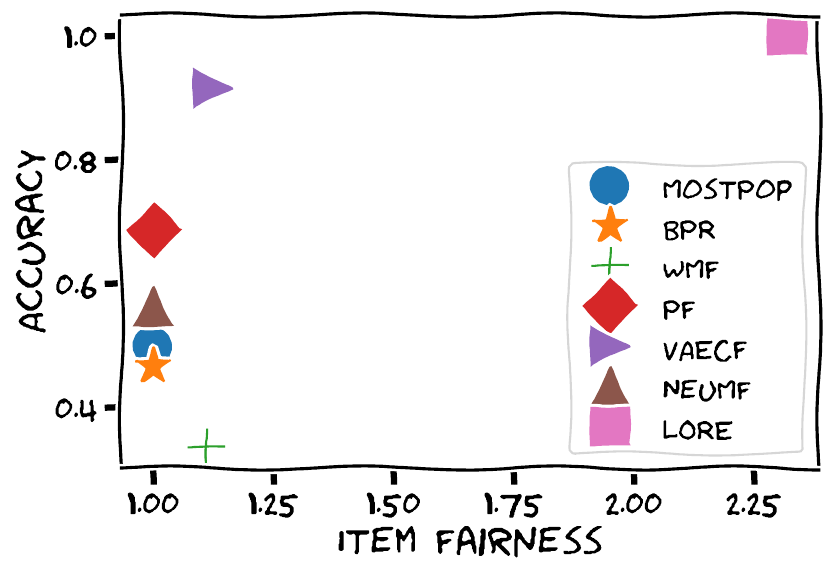}\label{fig:gowalla_ai}}
    \hfill
    \subfloat[Item \vs user fairness.]{\includegraphics[width=0.33\textwidth]{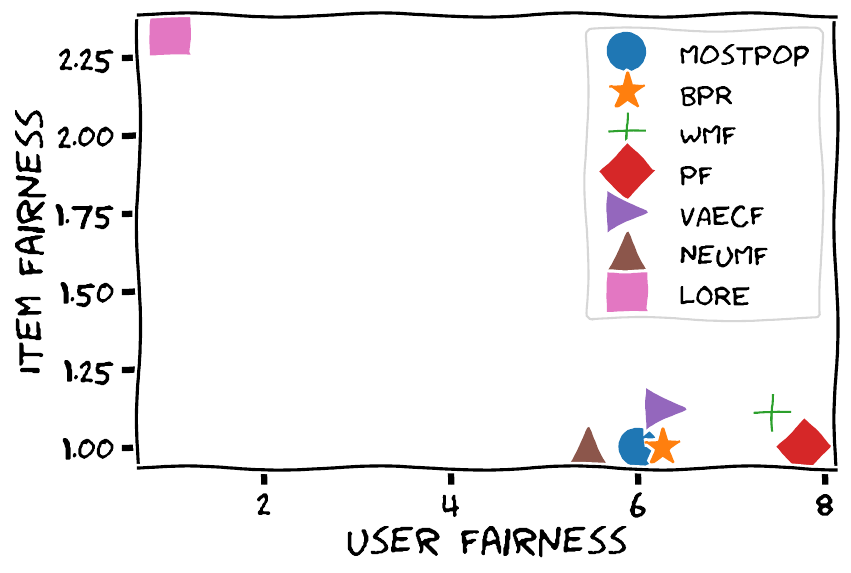}\label{fig:gowalla_ui}}
  \caption{Analysis of trade-off between three factors of accuracy, user fairness and item fairness in recommendation algorithms. The upper row represent the \yelp dataset while the lower row \gow. Note that algorithm that stay in the top-right hand corner are the best.}
  \label{fig:analysis}
\end{figure}

\subsection{Trade-off on Accuracy, User and Item Fairness (RQ2)}
\label{sec:tradeoff_analysis}
As we observed in Section \ref{sec:profile_analysis}, there is a bias in the POI check-ins data from both users' and items' perspectives, \ie specific POIs are visited frequently while many others are visited by fewer users. On the other hand, some users visits different POIs frequently while most users rarely visit the same POIs. In this section, we analyze whether and how the bias in the data would affect different classes of recommendation models. As can be seen in Fig.~\ref{fig:analysis}, we analyzed the relationship between three evaluation objectives, \ie \textit{accuracy}, \textit{user fairness}, and \textit{item fairness} with three 2D plots. Hence, each sub-figure (plot) captures two dimensions among three factors. Please note that we use $\frac{1}{MADr}$ to denote fairness (instead of bias), implying that for all dimensions, higher means better.
Hence, models that appear on the top-right corner of the plots achieve the best quality of recommendations based on specific evaluation dimensions of each plot (\eg accuracy \vs user fairness). The following insights can be obtained by analyzing the results generated by different models.

The majority of algorithms do not reside in the top-right corner of the plots, implying the existence of a trade-off between the accuracy and fairness of recommendation models.  For instance, although \lore is the best model in terms of accuracy and item fairness (see \lore on the top-right corner in Fig.~\ref{fig:gowalla_ai}), it results in poor performance on the user fairness, as for both Fig.~\ref{fig:gowalla_au} and Fig.~\ref{fig:gowalla_ui}, \lore resides on the top-left side (\ie low user fairness). This indicates that \lore sacrifice user fairness to achieve a good level of item fairness and accuracy. We will discuss this in detail in the following sections based on the results in Table~\ref{tbl:results}. On the other hand, from Fig.~\ref{fig:yelp_au} we can see that \wmf and \vaecf achieve comparable performance on both accuracy and user fairness, but they suffer seriously from item fairness issues (see Figs.~\ref{fig:yelp_au} and \ref{fig:yelp_ai}). These results show that it is hard to achieve a significant accuracy with both user and item fairness, and models usually sacrifice one dimension. For example, in our results, we have seen \lore and \wmf achieve the best result on accuracy and items fairness and accuracy and user fairness, respectively. However, neither model could produce satisfactory results on one dimension (\lore on user fairness and \wmf on item fairness).

A plausible explanation for this trade-off between different evaluation objectives is data biases. For instance, in Fig.~\ref{fig:data_analysis} it was shown that (active) users tend to use popular items; thus models such as \lore that achieve significant item fairness recommend the most popular items in their recommendation list. While these models are suitable for active users, they hurt other user groups' experience of recommendations measured in terms of accuracy.

\begin{table*}[!htbp]
\captionsetup{font=scriptsize}
\caption{Accuracy, user fairness, item fairness on \gow and \yelp dataset. Note that Pf$_0$=[0.25, 0.25, 0.25, 0.25], Pf$_1$=[0.7, 0.1, 0.1, 0.1], Pf$_2$=[0.1, 0.7, 0.1, 0.1], Pf$_3$=[0.1, 0.1, 0.7, 0.1], and Pf$_4$=[0.1, 0.1, 0.1, 0.7] characterize the fair distribution as uniform or non-uniform distribution (of resources) among four user groups, \ie very inactive, slightly inactive, slightly active, very active. Also, Pf$_0$=[0.33, 0.33, 0.33], Pf$_1$=[0.7, 0.15, 0.15], Pf$_2$=[0.15, 0.7, 0.15], and Pf$_3$=[0.15, 0.15, 0.7] characterize the fair distribution as uniform or non-uniform distribution (of resources) among item groups, \ie short-head, mid-tail, and long-tail. AUC represents the area under-curve for any of the evaluation objectives (accuracy, user fairness and item fairness).}
\begin{adjustbox}{width=1.15\columnwidth,center}
\centering
\label{tbl:results}
\begin{tabular}{lcccccccccccccccccc}
\toprule
\multirow{2}{*}{\textbf{Model}} & \textbf{Acc} && \multicolumn{6}{c}{\textbf{User Fairness}} & & \multicolumn{5}{c}{\textbf{Item Fairness}} && \multicolumn{3}{c}{\textbf{AUC}}\\
\cmidrule{2-2} \cmidrule{4-9} \cmidrule{11-15} \cmidrule{17-19}
& NDCG && Pf$_0$ & Pf$_1$ & Pf$_2$ & Pf$_3$ & Pf$_4$ & 1/MADr && Pf$_0$ & Pf$_1$ & Pf$_2$ & Pf$_3$ & 1/MADr && AUC$_{au}$ & AUC$_{ai}$ & AUC$_{ui}$ \\
\midrule
\multicolumn{19}{c}{Gowalla} \\ 
\midrule
\topp  & 0.0247 && \blue{-0.0246} & -0.6051 & -0.6039 & -0.6145 & -0.2985 & 6.0141 && -1 & \blue{-0.2143} & -2.8333 & -2.8333     & 1.0    && 1.497 & 0.248 & 3.007 \\
\bpr   & 0.0231 && \blue{-0.0255} & -0.6032 & -0.6135 & -0.6161 & -0.2958 & 6.2659 && -1 & \blue{-0.2143} & -2.8333 & -2.8333     & 1.0    && 1.459 & 0.232 & 3.132 \\
\wmf   & 0.0166 && \blue{-0.0228} & -0.3478 & -0.4917 & -0.6218 & -0.6461 & \underline{7.44}   && -0.7339 & \blue{-0.112} & -2.2175 & -2.2419 & 1.1108 && 1.245 & 0.185 & \textbf{4.132} \\
\pf    & 0.034 && \blue{-0.005}  & -0.4279 & -0.5023 & -0.5702 & -0.4674 & \textbf{7.7823} && -0.9928 & \blue{-0.210} & -2.8174 & -2.8174 & 1.0024 && \underline{2.667} & 0.343 & \underline{3.900} \\ \hdashline
\vaecf & \underline{0.0454} && \blue{-0.0041} & -0.4205 & -0.5052 & -0.5559 & -0.4795 & 6.3031 && -0.7453 & \blue{-0.108} & -2.2532 & -2.2664 & \underline{1.1244} && \textbf{2.884} & \underline{0.514} & 3.543 \\
\neumf & 0.0279 && \blue{-0.0226} & -0.5952 & -0.5943 & -0.6107 & -0.3057 & 5.4689 && -1 & \blue{-0.2143} & -2.8333 & -2.8333     & 1.0    && 1.538 & 0.281 & 2.734 \\ \hdashline
\lore  & \textbf{0.0496} && -0.1626 & -1.0985 & -1.0541 & -0.8962 & \blue{-0.157}  & 1.0    && -0.1682 & \blue{-0.0248} & -0.9053 & -0.858 & \textbf{2.3196} && 0.5   & \textbf{1.159} & 1.159 \\
\midrule
\multicolumn{19}{c}{Yelp} \\ 
\midrule
\topp   & 0.0097 && -0.1514 & -1.0666 & -1.0083 & -0.8917 & \blue{-0.1512} & 3.887 && -1      & \blue{-0.2143} & -2.8333 & -2.8333 & 1.0   && 0.488 & 0.138 & 1.943 \\
\bpr    & 0.009 && -0.1647 & -1.1049 & -1.0495 & -0.9236 & \blue{-0.145}  & 3.962 && -1      & \blue{-0.2143} & -2.8333 & -2.8333 & 1.0   && 0.497 & 0.138 & 1.981 \\
\wmf    & 0.0204 && \blue{-0.0196} & -0.6222 & -0.6047 & -0.5197 & -0.3359 & \textbf{6.119} && -0.5656 & \blue{-0.0626} & -1.8147 & -1.8661 & 1.245 && \textbf{2.049} & 0.459 & \textbf{3.810} \\
\pf     & \underline{0.0327} && \blue{-0.0294} & -0.6497 & -0.644  & -0.5711 & -0.2949 & 3.241 && -0.8386 & \blue{-0.1436} & -2.4688 & -2.4746 & 1.069 && 1.400 & 0.509 & 1.733 \\ \hdashline
\vaecf  & \textbf{0.0334} && \blue{-0.0166} & -0.5794 & -0.5982 & -0.5494 & -0.3322 & \underline{4.468} && -0.9292 & \blue{-0.1815} & -2.6753 & -2.676  & 1.031 && \underline{2.027} & \underline{0.515} & \underline{2.304} \\
\neumf  & 0.0086 && -0.1671 & -1.1196 & -1.058  & -0.8992 & \blue{-0.1646} & 3.943 && -1      & \blue{-0.2143} & -2.8333 & -2.8333 & 1.0   && 0.495 & 0.138 & 1.971 \\ \hdashline
\goso   & 0.0197 && \blue{-0.2782} & -1.4354 & -1.3774 & -1.084  & -0.2174 & 1.101 && -0.1009 & \blue{-0.0574} & -0.6934 & -0.7062 & \textbf{3.096} && 0.184 & \textbf{0.571} & 1.706 \\
\lore   & 0.0197 && \blue{-0.2583} & -1.385  & -1.3258 & -0.9952 & -0.2519 & 1.148 && -0.1368 & \blue{-0.0374} & -0.8192 & -0.7769 & \underline{2.580} && 0.194 & 0.480 & 1.481 \\
\bottomrule
\end{tabular}
\end{adjustbox}
\end{table*}

\subsection{Popularity Bias in POI Recommendation (RQ3)}
\label{sec:biasrecommendation}
A comparison between recommender algorithms can be made by comparing the accuracy measured by NDCG, user fairness, and item fairness values in Table~\ref{tbl:results}. We utilized two metrics in measuring fairness, GCE and MADr. In addition, we computed the area-under-curve (AUC) for the 2D plots shown in Fig.~\ref{fig:analysis}.
Due to space constraints, we only show results of models computed based on NDCG$@10$.\footnote{The results of other models and metrics are available at \url{https://recsys-lab.github.io/FairPOI/}} Note that the largest NDCG corresponds to the most accurate system, while the lower GCE (in an absolute sense) corresponds to the fairest system (for equal distribution GCE=0). 
While GCE can tell toward which group the system is providing the highest quality of recommendation or all groups receive relatively similar quality (\ie fair recommendation), MADr can only demonstrate the latter. We compute the AUC according to Fig.~\ref{fig:analysis} for each three relevant factors on both datasets. We use \textbf{bold} and \underline{underline} to show the best and second best results in NDCG, $\frac{1}{MADr}$, and AUCs. Also, we use the \blue{blue} color to highlight each recommendation model provides the best quality of recommendation to different groups of users, \ie \textit{very inactive}, \textit{slightly inactive}, \textit{slightly active}, and \textit{very active} in user groups and \textit{short-head}, \textit{mid-tail}, and \textit{long-tail} for item groups. For instance, in the user fairness, if the \blue{blue} color (minimum value of the GCE) occurs under pf$_0 = [0.25, 0.25, 0.25, 0.25]$, the recommendation model produces a fair recommendation, while if this happens under pf$_1 = [0.7, 0.1, 0.1, 0.1]$ the algorithm is providing better quality to very inactive users. Thus, we can identify the tendency of an algorithm to provide more fair (GCE is minimum under $pf_0$ and MADr is minimum) or less fair recommendations in favor of/against a particular group of users based on their activity level. 
By analyzing the results presented in Table \ref{tbl:results}, the following insights can be identified.

The results on \gow and \yelp datasets show different patterns for accuracy and fairness (\ie users and items). Among all categories of models, the neural approaches (\ie \vaecf and \neumf) overall achieve the best results among other models on both datasets. For instance, \vaecf achieves the best and second-best performance in terms of the accuracy (NDCG) in \yelp and \gow datasets respectively. On the other hand, \lore, which is a contextual model, had the best performance on \gow, while \yelp's accuracy was not significant. Moreover, \vaecf can produce fairest recommendation \wrt both user and item fairness. The results on $\frac{1}{MADr}$ on both user and item fairness show that \vaecf could perform well in comparison with the best and second-best models. However, the AUCs' results on both \gow and \yelp datasets indicate that \vaecf is the best and second-best approach on all three relevant factors. These results are proof of the overall performance of \vaecf, \ie while \vaecf can achieve significant accuracy, it also can produce an almost fair recommendation based on both consumer and provider sides.

Among the traditional models (\ie \topp, \bpr, \wmf, and \pf), the result of \bpr is slightly surprising as it very is similar to the \topp. Our explanation for this observation is that due to the sampling we performed on the dataset for speeding up the experiments as performed in~\cite{kowald2020unfairness}, our dataset has been more skewed toward popular items. This can hamper the performance of BPR, which has been shown to be sensitive to popular items~\cite{deldjoo2021explaining}. More importantly, although \wmf achieves much better results than the \topp and \bpr, in most cases \pf has the best performance among the traditional models. This indicate that \pf using the Poisson distribution can capture the long-tailed user activity found in most consumption data and model the relation between user and items on implicit feedback (\ie check-ins) in a better way. This leads to better estimates of users' latent preferences, and therefore superior recommendations, compared to competing methods. However, \yelp tends to produce unfair recommendations for user groups (\ie GCE minimum under $pf_4$ under user fairness), while on the \gow dataset \lore recommends most popular items that is interesting for active users. 

The observation on item fairness shows that all methods have the lowest value of GCE (in absolute sense) under $p_{f_1}$ --- \ie popular items receive higher exposure opportunities. This pattern for the results for item fairness can be verified on both datasets, \ie \gow and \yelp. Thus, we can note a clear unfairness of popularity bias that is introduced in all methods. However, between all models contextual based models (\ie \lore and \goso) overall improve item fairness (\ie maximum $\frac{1}{MADr}$) on both datasets. This indicates that the impact of contextual influence on POI recommendation can provide diversity of items in the recommendation lists.

\section{Conclusion and Future Work}
\label{sec:conclusion}
It is crucial to avoid potential data or algorithmic bias, respecting the primary task of proposing personalized recommendations in POI. 
We analyzed the trade-off between accuracy, active users, and popular item exposures fairness where advantaged and disadvantaged user and item groups are determined based on the interactions. Experimental results show that numerous methods failed to balance the item and user fairness due to the natural biases in data. Among 8 POI recommendation models (of different types traditional CF, neural, and contextual), only \vaecf produces the best level of accuracy, user fairness, and item fairness. Essentially this shows that \vaecf is learning the underlying user-item representation accurately and in the least unbiased way. This result is interesting and confirms the results reported in \cite{dacrema2019we}, where between all neural models, only \vaecf learns the best underlying representation over the complex non-neural models. Our analysis confirmed this performance also on fairness dimensions. Also, there is variability across algorithms' performances across two datasets. Usually, one aspect, item fairness or user fairness, is compromised to keep the accuracy high.
Moreover, we observe that a primary challenge many models face is the unfairness of popularity bias.
As for future work, we plan to extend our experiments on more datasets from different domains (e.g., marking in e-commerce \cite{wan2020addressing}, media-streaming, e-fashion \cite{deldjoo2022review}) and models such as session-based \cite{quadrana2017personalizing}, neural \cite{dacrema2019we} and content-based systems \cite{deldjoo2018content}. Investigating the different type of bias, such as gender bias, can be a potential direction for future research. Moreover, proposing a single metric that can serve for the evaluation of all these dimensions for a given recommendation algorithm is useful to provide a way to directly compare different recommendation methods considering various types of biases without complicated analysis.
We also plan to study approaches that can mitigate two-side fairness while producing accurate recommendations. We also deem the study of multi-sided fairness from a security perceptive an interesting open direction \cite{deldjoo2020adversarial,deldjoo2021survey}. Finally, we plan to examine the impact of data characteristics~\cite{deldjoo2020dataset,sachdeva2022sampling} and hyper-parameter tuning on all the above-reported dimensions.

%
%
%
\bibliographystyle{splncs04}
\bibliography{mybib}
\end{document}